# Ad Hoc HLA Simulation Model Derived From a Model-Based Traffic Scenario

David Reiher
*Carl von Ossietzky University of Oldenburg*
Oldenburg, Lower Saxony, Germany
david.reiher@uol.de

Axel Hahn
*German Aerospace Center (DLR) SE*
*Carl von Ossietzky University of Oldenburg*
Oldenburg, Lower Saxony, Germany
axel.hahn@[dlr,uol].de

*Abstract*— **Modern highly automated and autonomous traffic systems and subsystems require new approaches to test their functional safety in the context of validation and verification. One approach that has taken a leading role in current research is scenario-based testing. For various reasons, simulation is considered to be the most practicable solution for a wide range of test scenarios. However, this is where many existing simulation systems in research reach their limits. In order to be able to integrate the widest possible range of systems to be tested into the simulation, the use of co-simulation has proven to be particularly useful. In this work, the High Level Architecture defined in the IEEE 1516-2010 standard is specifically addressed and a concept is developed that establishes the foundation for the feasible use of scenario-based distributed co-simulation on its basis. The main challenge identified and addressed is the resolution of the double-sided dependency between scenario and simulation models. The solution was to fully automate the generation and instantiation of the simulation environment on the basis of a scenario instance. Finally, the developed concept was implemented as a prototype and the resulting process for its use is presented here using an example scenario. Based on the experience gained during the creation of the concept and the prototype, the next steps for future work are outlined in conclusion.**

*Keywords—High Level Architecture, traffic simulation, traffic scenario, scenario-based testing, data modeling, V&V*

## I. Introduction

The development of automated and autonomous vehicles has made significant progress in recent years. Nevertheless, we are still a long way from the widespread introduction of autonomous vehicles [1]. Some traffic domains are more in the focus of the general public and are therefore more technically advanced, but efforts to reach the goal of fully autonomous vehicles suitable for widespread everyday use unite all of them. A certain historical and cultural pioneering role can be attributed to road transport systems[2], but water transport systems[3,4] and other domains' transport systems are increasingly experiencing great interest in this regard, too.

One of the biggest challenges in developing such autonomous vehicles (AVs) is their validation and verification (V&V). Especially for the successful introduction of future vessels with a degree of autonomy of three or higher, as categorized by the International Maritime Organization (IMO)[5], and future road vehicles with driving automation of level of 3 or higher as defined by SAE[6], the proof of their functional safety is very important: Studies have shown that a large part of the population is concerned about AVs malfunctioning and therefore lacks the necessary level of trust to use them in the near future[7–11]. From this, it can be concluded, as a study of the German and U.S. print media[12] also showed, that one of the most important aspects on the way to public acceptance of highly automated and autonomous vehicles is the clear proof of their functional safety.

Due to the introduction of non-deterministic approaches, such as the use of self-learning artificial intelligence, classical deterministic functional safety verification methods such as model checking and theorem proving alone are often no longer sufficient to ensure the functional safety of the whole system or its parts.[13,14] Instead, the behavior of these systems must be checked at least partially at runtime. Unfortunately, it is not possible to manually test all conceivable situations an automated or autonomous vehicle could be exposed to in the future with reasonable effort. Covering all permutations of environmental variables by real-world test drives requires an economically unfeasible amount of time and resources[15]. The last few years have brought along some approaches to tackle this issue. A very popular approach to replace real-world testing is the use of traffic simulations.[16] Running tests in a simulated virtual environment brings several advantages, including the ability to be able to be used during development, to be executed faster than in real-time, and not to expose people and equipment to any risk[17]. Especially the use of simulation for validation of sub-systems as early as during the development stages can lead to early and cost-efficient detection of errors.[17] In order to obtain meaningful results from simulative tests, an additional systematic approach is required.[18] For this purpose, the so-called scenario-based approach, as proposed within the Pegasus project[19] and others, has become established. The approach consists of identifying, modeling, simulating, and evaluating exactly those traffic scenarios that are most relevant for obtaining meaningful information about the functional safety of a particular (sub-)system under test (SuT).

The need to be able to test subsystems of a vehicle in a simulative way results in the necessity of the possibility to combine simulations from different providers or even to replace components of a simulated vehicle with one or more real systems, external simulations, external pieces of software, or external models. For this purpose, co-simulations have proven to be particularly useful.[20–27] This is mainly because their distributed nature allows for external systems to be connected by design. Distributed Co-simulations, however, bring their own challenges. For example, the data that is exchanged between the participating simulations must be additionally described and this description must be distributed among the participants so that they know what kind of data in what format is being sent by the others and how to deal with it. Some co-simulation standards also introduce a central communication interface that needs to know which participant wants to receive which data. At first glance, this approach is not necessarily compatible with the scenario-based approach since different scenarios with different requirements are to be simulated in short succession.

## II. CO-SIMULATION STANDARDS

In addition to many concrete use case, tool, or technology-specific co-simulation frameworks that can be found in the current scientific literature, two standards have emerged that are by now widely used. The Functional Mock-Up Interface (FMI) and the High Level Architecture (HLA) standards have a very similar goal, are both described as tool independent, but have some significant differences in approach on closer inspection.

The Functional Mock-Up Interface (FMI) is a standard for the exchange of dynamic models and co-simulation[28]. The first version of the standard was published in 2010. Since 2011, the maintenance and further development of the standard have been carried out by the Modelica Association, a second version of the standard was published in 2014 and the third version is currently under active development. FMI for co-simulation establishes a co-simulation environment where multiple simulation components are coupled. Those components implementing the FMI standard are called Functional Mock-up Units (FMU). On a basic level, an FMU can be described as a wrapper of an XML file, a solver, and the model to be executed deployed as C source code or a platform-dependent binary file. To run a co-simulation consisting of several FMUs, a master algorithm is needed. A corresponding ready-to-use implementation is not provided by the standard, but there are some free and commercial ones. These differ in the supported functions because the FMI standard does not prescribe a faster set, interfaces, or processes that have to be fulfilled by the master algorithms implementation. Thus, only some of the master algorithms support, for example, a physically distributed operation. The standard has been developed for efficient simulation of continuous systems, whereby the data exchange between the FMUs, however, only takes place at discrete points in time.

The High Level Architecture (HLA)[29] standard was originally defined in the 1990s by the Defense Modeling and Simulation Office (DMSO) of the US Department of Defense (DoD) to facilitate the assembly of stand-alone simulations. Since 2000 it is a standardized, regulated, and published by IEEE. The original goal was the reuse and the interoperability of applications that were not necessarily interoperable by design. Therefore, HLA is meant to resolve interoperability and reusability issues between those software components. Another important aspect of the HLA specification is the synchronization capability. HLA can be used for implementing discrete-time simulations as well as for discrete-event simulations, or a mixture of both. Much emphasis is therefore placed on time management and synchronizing the participating simulations, called federates, to avoid inconsistent states of the overall simulation, called federation. To achieve this goal, there is a central component called Runtime Infrastructure (RTI). It handles tasks such as federation management, sharing of objects and their values between federates, time management, etc. In addition, federates do not communicate directly with each other, but only indirectly through the RTI, as can be seen in Figure 1. To do this, the RTI needs to know what kind of data each federate will provide and what it wants to consume. These data and relationships are described in the so-called Federation Object Model (FOM). This in turn can be extended by each federate in the sense of so-called modular FOMs since the introduction of HLA Evolved (IEEE 1516-2010) in 2010.[30] HLA only describes interfaces, processes, and contracts to be complied with, but does not provide a ready to use implementation of the RTI. However, thanks to the active community, there are some free and commercial implementations available.

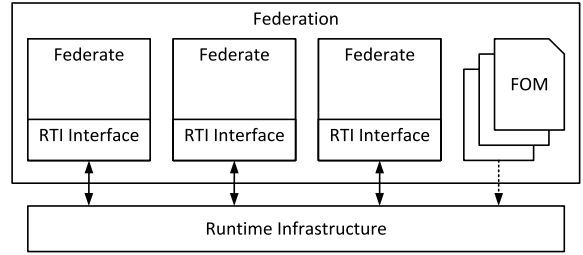

*Figure 1: The basic structure of an HLA-compliant co-simulation.*

One of the main differences is that one could arguably say that a single FMU in an FMI co-simulation can be considered as a replacement of a real sub-system ready to take inputs and compute resulting outputs. This is closely related to the definition of a digital twin of a single traffic participating vehicle.[31] FMI is, therefore, more suitable for representing a vehicle in detail as a system of systems. HLA, on the other hand, places emphasis on a single federate representing larger entities or even groups of entities. This fits very well with the structure of an agent-based simulation, which is very close to real traffic if one considers individual vehicles or their drivers as agents. In addition, the HLA standard places more emphasis on supporting physically distributed simulation environments, which has great added value when the system under test is located elsewhere and can only be connected via communication infrastructures. These two mentioned advantages, together with the very detailed prescribed interfaces, workflows, and constraints, as well as the focus on time and data management, make HLA the more suitable choice for the further proceeding of this work.

### III. HLA FOR FLEXIBLE SCENARIO-BASED TEST WORKFLOWS

As mentioned in the previous section, the federates of an HLA federation do not communicate directly with each other but indirectly using the central RTI. A federate informs the RTI when it instantiates new simulation-relevant objects and sends the corresponding values to the RTI each time the attributes of these objects are updated. Analogous to this publishing mechanism, a Federate informs the RTI which object instances and attributes of other Federates it is interested in and subsequently receives the new values from the RTI after each corresponding attribute update. For this to work generically - to ensure interoperability - the data is sent to and received by the RTI. For this to work, the structure of the objects and the data types of their attributes have to be defined manually in the Federate Object Model (FOM).

The development of HLA federates is typically very complex and resource-intensive because developers must invest a significant amount of time not only to create and maintain the just mentioned FOM but also to handle common HLA functions such as controlling simulation time, the synchronization process between federates, publishing, subscribing, and updating elements of the objects along with their associated coders and decoders. As a result, they cannot fully focus on the actual functionalities of the simulation content.[32] While the functional aspects, such as time synchronization and communication with the RTI, can be fairly easily moved to a library since these processes are similar enough for most federates, the FOM must always be adapted to the respective communicational functionality of an entity to be simulated. The methods for processing the received object instance and attribute updates must also be adapted accordingly to the behaviors mapped by the federate implementation.

This approach is well suited for purposes where one has a lot of different stand-alone simulators that rarely if ever get customized and get combined in a plug-and-play manner depending on the planned simulation content. This is for example the case for vehicular and military training and education simulation environments, which are a common application for HLA. However, this poses a challenge for the efficient use of HLA for simulative scenario-based test workflows. In this case, the simulation contents, their level of abstraction and their behavior are often adjusted, since a wide variety of scenarios must be simulated, with sometimes more and sometimes less complex participants, behaviors, and environments. This is due to the fact that the simulation content and flow must be tailored to the system under test and its current development status.[17] In addition, scenario-based test runs often consider extreme and rarely occurring situations, which may require unusual and arbitrary behaviors or parameter combinations. These can also vary from scenario to scenario.

On the other hand, predefined FOMs and communication implementations are advantageous for the reuse of implemented federates and for long-running simulations to which new sub-simulations are added during runtime. Reusability is also of utmost importance for the scenario-based approach, as it is essential for its efficient use that not every scenario has to be completely implemented from scratch. However, adding sub-simulations at runtime is not relevant in this case and can be discarded as a requirement, since the ideal scenario-based approach envisions that many different scenarios are simulated in sequence and that they are self-contained. Instead, distributed co-simulation is mainly used here, as already touched upon, because it allows an intuitive connection of the vehicle as the system under test, in the sense of vehicle, hardware, software, or model-in-the-loop.[17] In addition, this provides good load distribution capabilities for the simulation of more complex scenarios.

## IV. INTERDEPENDENCE BETWEEN SCENARIO AND SIMULATION MODELS

Since scenarios and their representation are, as the name suggests, the core of scenario-based testing, they have a key role to fulfill. This is especially true when scenario-based testing is integrated into the development process to validate the results of each development step, as described in a previous work on a new approach to describing traffic scenarios with a model-based multilayered approach.[17] This development integrated process is illustrated in Figure 2.

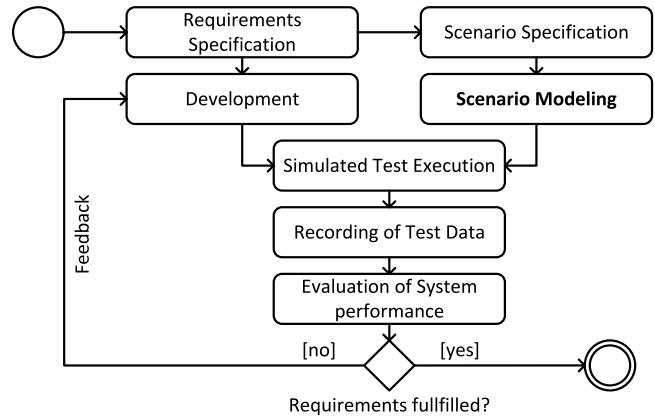

*Figure 2: Activity diagram describing the scenario- and simulation-based Validation and Verification process of automated and autonomous traffic systems as part of the development phase.[23]*

The classical approach to initiating simulation runs would be that a simulation system does contain a simulation model (the implementations of the simulated entities and their dependencies), which represent the simulation run's contents, and these are set into a certain starting state via direct parameterization, from where the simulation run is started. Scenario description languages (SDLs) in particular have established themselves for this task. There are many of them, some open, others closed, some text-based, others graphical. Two examples are OpenSCENARIO[33], which is very well established in the automotive field and represents the contents of a simulation run in a text-based way, and Traffic Sequence Charts (TSCs)[34], which allow the graphical representation of trajectory families based on formal semantics. Most of the established SDLs are strongly bound to a specific simulation model or have to be adapted to it if the scenario is to be executed by a simulation system. A good overview of established SDLs in the automotive sector can be found in the publication of Ma, Che et al.[35]

Due to the need for reusability and persistent cataloging, when a simulation is used during scenario-based testing, this initial state - simply put - is now separated from the state of the simulation and is referred to as a scenario model. The separation leads to two independent models, which are interrelated. This in turn leads to some challenges in modeling and simulating such scenarios, which are largely similar to the difficulties described in the previous section. The resulting potentially problematic areas are shown in Figure 3. Since the set of possible scenario content has to represent the possible contents of a simulation run one-to-one in order to maintain compatibility and utilize the full potential of the simulation system,[17] there is a certain interdependence between the simulation and scenario models. This interdependence means that when one model is modified, the other model must also be adapted. Looking more closely at the two models, it becomes apparent that, in the worst case, the simulation developer has to modify four artifacts in total for every simulation run that is to have new or altered simulating capabilities: The implementations of the simulation objects, the FOM, which describes which data is exchanged within the simulation, the objects for scenario description, and the scenario description in terms of configurations and parameterizations. As a result, this kind of scenario-based use of simulations is not practically feasible.

A possible solution to this problem was presented in a previously published work.[17] The core idea is that the simulation models are generated based on the scenario models without the need for manual adjustments. Thus, the interdependence would be resolved or, strictly speaking, obscured for the simulation developers and users. In short, a multi-level model-based approach has been proposed that makes it possible to create scenarios based on a library of scenario building blocks and to use these as direct input for an otherwise rudimentary simulation system. These modelling libraries themselves consist of a model based on a UML extension developed for this purpose and its implementation in the form of Java code. Since this work was more conceptual in character, the present paper will deal with the fundamental technical setup necessary for its utilization in the context of an HLA-based distributed co-simulation. The result is a framework that allows the development of scenarios to be simulated for different traffic domains without the need for HLA-specific knowledge. In addition, the interdependence of the simulation and scenario models has been resolved, resulting in a more streamlined scenario development process. This in conjunction leads to the effect that not only simulation engineers can use simulation systems, but the developers of the systems to be tested can ideally directly create and simulate the necessary scenarios, which allows the process shown in Figure 2 to be implemented more efficiently during the development of (highly-) automated systems.

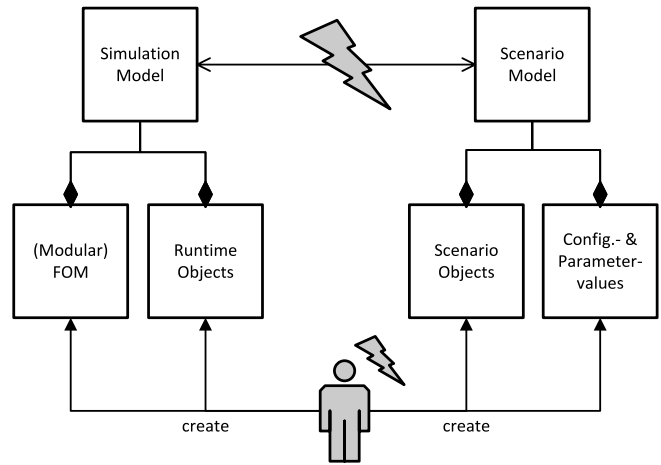

*Figure 3: Illustration of the tension between simulation and scenario models in the context of HLA-based simulations.*

## V. Related Work

After the publication of the first HLA standard and the first projects using it, it was quickly realized that the development of an HLA-compliant federation is very complex. Therefore, as early as the late 90s,[36] efforts were made to simplify the development and increase the reusability of simulation components. Initially, the main focus was on the reusability of FOMs to avoid having to rewrite them over and over again. Modular concepts were additionally developed in order to be able to reuse separate simulation components more easily. In the later years then some approaches were developed, with which the HLA specific functionality can be hidden partially or completely from the developer, to avoid the repeated creation of infrastructural boilerplate code. This should make it easier for developers to start using HLA and reduce the time required to implement a running federation.

For the objectives of this work as described in the previous section, the reusability of model components, the automatic generation of HLA-specific components, and other development easing techniques are also possible relevant concepts and assessed for their suitability.

### A. *Base Object Models*

The idea behind Base Object Models (BOMs) is to provide a component framework for facilitating interoperability, reuse, and composability, to tackle the problems caused by the increased complexity of simulation environments. The concept is based on the assumption that piece parts of models, federates, and federations can be extracted and reused as modeling building blocks or components.[37] The concept of BOMs was first introduced in 1998,[36] and later in 2006 became a standard maintained and published by the Simulation Interoperability Standards Organization (SISO).[38]

Essentially, a BOM serves to represent a component. The focus is placed on describing the interface for a component, not the implementation details. It is therefore the responsibility of the simulation system to provide the implementation of behaviors described by the interface.[39] This description is persisted in the form of static data structures such as tables, UML, and XML and can thus be stored and indexed in a kind of library for reusable co-simulation participants. A developer then searches this library for a BOM that meets previously identified requirements for Model Capacities. If such a BOM exists, it can be used instead of developing a new one. This procedure is thus located in steps 2 to 4 of the Federation Development and Execution Process (FEDEP).[40] The behavior described by the selected BOMs can then be implemented manually into the federate or an already available BOM Component Implementation (BCI) can be selected and integrated into the federate. The idea behind BCIs is to provide component model implementations matching the required behavior described by a component interface, to increase the reusability even further.

Although this approach originally brought some distinct advantages, such as significantly increasing the reusability of simulation components, increasing comprehensibility, and reducing complexity for developers, BOMs alone are not a suitable approach for implementing the short-lived scenario-based co-simulation runs envisioned here. Since BOMs represent structures and processes separately from the implementation and scenario description, the problem of interdependencies identified earlier, and the associated maintenance effort remain. A catalog like the use of BOMs and BCIs provide is very well suited for simulation applications where components are rarely tweaked and instead often recombined. For the use case envisioned here of using scenarios as a direct input for initializing the simulation environment, the use of BOMs would actually create yet another third dependency: structure and communication model, implementation of federates and/or BCIs, and the scenario.

## B. BOM Modeling Framework (BMF)

In 2011 a BOM-based framework prototype that supports model editing, code auto-generating, testing, and component-based modeling, called BOM Modeling Framework (BMF), was presented. The goal of this framework is to further promote the reusability and interoperability of models, in addition, to further decrease the development complexity of HLA-based co-simulations.[41] This is done by further partitioning the executable part of a federate (see Figure 4): Atomic Models are the central building blocks in this approach. They are executable small units that provide certain functionalities and can be combined with other Atomic Blocks to form a Coupled Block. Each Atomic Block has precisely defined inputs and outputs. The Atomic Block itself is also subdivided into the so-called Kernel Model and the Connected Model. The kernel model contains the models and implementations required for the intended domain-oriented functionality. The kernel model contains all the necessary models and infrastructural implementations that are required for communication and interaction with other blocks. This results in a clear separation between the business functionality and the structural models and implementations: "Any model must provide an abstract interface in which various operations are defined. Concrete implement it. User-defined models never one must inherit directly invoke methods of other user-defined models. Models must never depend on specific implementations of other models."[41] Another important part of the BMF is the Extensible Simulation Running Framework (XSRFrame), which ultimately ensures that the models are able to be executed and interact with each other. Actually, XSRFrame is an HLA-compatible general-purpose federate that can accommodate BOM models and provides the most important interfaces and services to communicate with the RTI.

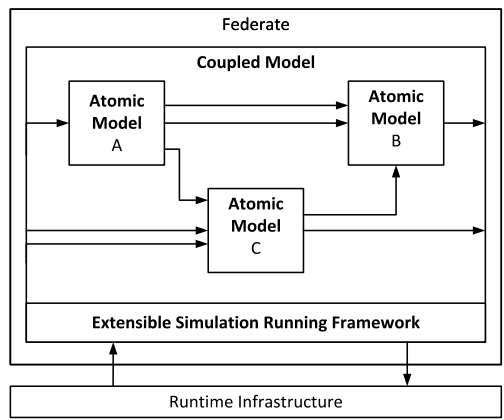

*Figure 4: Exemplary basic structure of a BMF based federate.[41]*

The general procedure for development with the BMF is that a model description document is created first. Based on this, skeleton code is generated automatically, which must be implemented by the developer manually to the end. The code and the model are then exported as an atomic model in the form of a DLL. Tool support is required for these steps and the subsequent generation of the Connected Model functionalities. This creation of Coupled Models from multiple Atomic Models can then be done manually. Although the use of the BMF greatly increases the reusability of models and implementations and additionally reduces the complexity for the developer, the dependency between models is at most somewhat obscured. In the case of more heavyweight modifications, the model description must first be modified again in order to generate the corresponding code, which then in turn has to be modified too. In addition, the publications on the BMF do not deal with the question, in which extent the parameters of the models can be influenced from outside of the DLLs themselves. The latter, however, is of great importance for a scenario-based approach. On the other hand, the strict separation between interfaces and implementation as well as between model and engine is a step in the desired direction of scenario-based simulation environment instantiations and should therefore be given continued attention in the further course.

## C. HLA Development Kit Software Framework

The HLA Development Kit software Framework (DKF) is a general-purpose, domain-independent, open-source framework, which facilitates the development of HLA Federates. The DKF allows developers to focus on the specific aspects of their own federates rather than dealing with the implementation of the common HLA-specific functionalities like managing the simulation time, connecting to the RTI, publishing and subscribing, and managing the HLA specific Object and Attribute elements.[42] The DKF is built around three key principles: (1) Interoperability, which is achieved through full conformity to the IEEE 1516-2010 standard specifications; (2) Portability and Uniformity, DKF provides a homogeneous set of APIs that are independent of the underlying HLA RTI and Java version; (3) Usability, the complexity of the features provided by the DKF framework are hidden behind a set of APIs.[43] The basic structure of a DKF-based federation can be seen in Figure 5. The DKF layer is also able to accept extensions and use them for application-specific processing tasks. Thus, the framework can also be extended to a certain extent for specific applications.

The DKF's approach successfully separates HLA functionalities and the function calls required for them from the implementation of the actual federation. As a result, the need for expert knowledge of HLA functionalities is reduced and the development of small test/dummy federations as well as large real federates is accelerated. Its clearly defined structures, interfaces, and processes also make it easy to understand the internals of the processes within an HLA federation and inside the DKF layer. Unfortunately, the main drawback of the two-way dependency between the models cannot be completely solved here either: FOMs still have to be written by hand and the corresponding Java classes subsequently have to be annotated with the DKF's built-in annotations. A scenario-based approach is also not possible in a straightforward way when utilizing the DKF, since a scenario - as understood here as a combination of simulation participants and their starting states - cannot be used as a direct input for the simulation environment's initialization. Nevertheless, the DKF approach comes close to reaching the goal set for this work and some concepts can be used as guidance, such as the unified API for implementing a federates internal behavior.

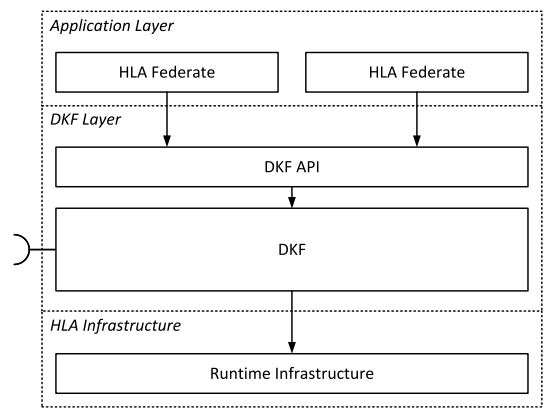

*Figure 5: The basic architecture of a DKF-based Federation.[42]*

## D. Model-Driven Approaches to the Development of Distributed Simulations

Another approach that has received increasing attention over the last decade is the model-based generation of distributed simulations. Since this paper focuses on the use of HLA (cf. Section 2), only the most prominent approach in this area that uses HLA will be highlighted in the following.

In 2012, Bocciarelli, Pieroni, et al. proposed a model-driven method for building distributed simulations.[44] At this point, the work focused on the transformation of business process models into executable distributed simulations for the purpose of analysis and testing. Since transport systems differ from business processes in many ways, the work would not be relevant at this point. In the same year the basic idea was extended to simulate general system models by building upon the Systems Modeling Language (SysML).[45,46] The basic idea is to create a SysML model of the planned federation, consisting of several individual model instances, which are then enriched with HLA-specific details using HLA-specific UML profiles, and later transferred into an implementation by a multi-stage model transformation process. The semi-automatically generated implementations are merely skeletons that have to be filled with logic subsequently to get an executable federation. This largely avoids manual implementation of HLA-specific processes, which makes the development process of distributed federations less complex.

This approach was further expanded and improved in the following years by cloud-based deployment capabilities[47], tailoring the approach to the Object Management Group's Standard for Model Driven Engineering, the Model Driven Architecture,[48,49] and adding automated FOM generation to further ease the process of creating a full-fledged executable federation.[50] The last point in particular solves a major weakness of the previously mentioned related works: The manual creation of the FOM. The above-mentioned DKF was in fact used in one publication as the basis for the creation of the implementation.[49] It could therefore possibly be used for closing one of the biggest gaps of the DKF by placing the model-based process upstream.

The latest iteration is the tailoring to the standardized Distributed Simulation Engineering and Execution Process (DSEEP)[51], which emerged from the HLA standardization activities and provides a standardized and rigorous process for developing and executing distributed simulations. The result goes under the name of Model-Driven Distributed Simulation Development Process (MoDSEEP).[52] The applicability of this model-driven approach was later demonstrated on the basis of use cases from practice and concrete tool chains.[53,54]

Although this approach has very similar overarching goals to the work presented here, is obviously very well elaborated and seems to be practically applicable, the concrete objectives are slightly different, which renders the results highlighted above not directly applicable for fulfilling the goals pursued here. The problems identified are the following: (1) To perform the proposed model-based process, HLA-specific knowledge is still needed for the manual enrichment of the SysML models using the two proposed HLA-specific UML profiles. The approach presented here completely hides all HLA mechanisms from the user. (2) The implementation of the actual logic of the federates is downstream of the multi-level model transformations, as is common in MDA. Thus, when using scenarios as understood in this work, the integration of the logic would have to be done again for each scenario. The approach proposed here moves the implementation further upstream in the process, so that the scenario-creating user of the simulation system only has to assemble and parameterize ready-made "building blocks". (3) The proposed tool chains contain a not insignificant number of different applications and tools to perform the individual steps of the process. Proprietary software like Pitch Developer Studio[55] is also used in some cases. The goal of the present work is to have significantly fewer dependencies. (4) The resulting implementation does not seem to give the developer an easy way to use the data (objects and attributes) published by other federations. Or at least this question remains open, as the published papers do not address this in detail. The approach presented here offers the developer a uniform way to easily access objects and attributes from other federations and use them for own calculations within the federate logic as is demonstrated in the later course. Nevertheless, knowledge from the existing Model-Driven Approaches can and should be included, especially since the basic idea overlaps with the one building the foundation here.[17,23]

## VI. Ad Hoc Simulation Model

As already touched upon in section IV, this work aims to explore how to implement the technical foundation for a realization of a previously published conceptual work.[17] The presented approach is that the simulation model are generated directly based on the scenario model. This would resolve or, strictly speaking, obscure the interdependency for the simulation system developers and users. In addition, parts of the approaches are adopted from the related works mentioned in section **Error! Reference source not found.**: The developer should only have to interact with uniform and relatively simple interfaces in order to completely hide HLA's own quite complex functionalities and the need to implement the handling of HLA-specific management tasks every time. Also, as mentioned in section V.B, a strict separation between model and engine is aimed for, to allow simple modeling of the scenarios independent of any infrastructural functionality and independent of the HLA implementation used. In the following, the model-based approach with the mentioned additions obtained from the related works will be looked at on quite a high level to prepare for the following section on the actual implementation.

To resolve the double dependency between the models from the developer's point of view, the simulation model should be generated entirely from the scenario model. This is shown in Figure 6. Here it can be seen how the scenario model on the left side defines the basic structures, rules, and possible simulation components in a hierarchical way. The initial simulation model is then generated from the scenario instance created in this way. This can be thought of as a description of object-oriented program classes which, together with a set of parameters, are transformed into concrete object instances. As the simulation time progresses, these objects are updated in each time step, resulting in a new simulation state. This allows fast execution of different scenarios one after the other since no time-consuming adjustments have to be made to the actual simulation system before simulating a new scenario.

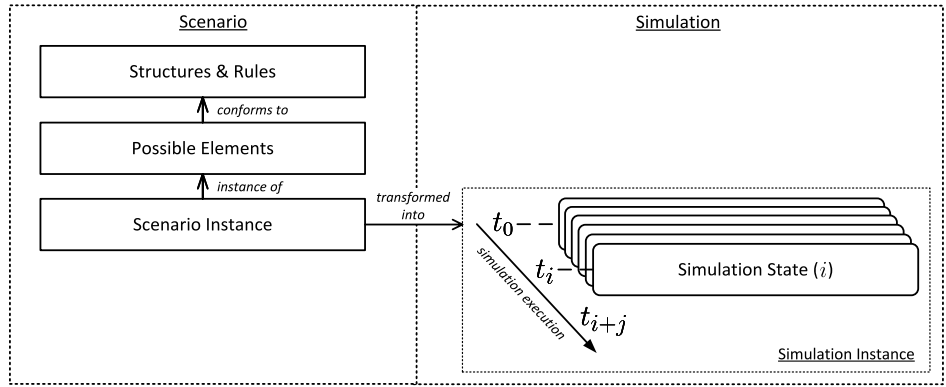

*Figure 6: Structure and dependencies of the scenario and simulation models (cf. previous work in Reiher & Hahn[17])*

To enable this design, the HLA-specific object models must also be generated from the scenario model. In more specific terms, this means that a valid FOM or valid FOM modules must be generated during the transformation. Together with the instantiated simulation objects, these form the simulation model. In order for simulation objects to be able to react to each other, the information about publishing and subscribing must also be included in the scenario and be utilized in the transformation. In order to provide the simulation system user with possible elements that can be used to compile and parameterize a scenario as simple as possible, a modular building-blocks-like approach is adopted. This can be seen in Figure 6 on the left side.

An intuitive implementation of this is a multi-layer object-oriented inheritance structure. Here, the functions, interfaces, and structures that are essential for the federation's functionalities are first defined by abstract classes. These are then inherited by classes that add additional domain-specific functions - here the maritime domain is used as an example from now on. These building blocks can then be assembled in the form of a scenario. This can be done in the form of a simple data structure such as XML, in which the selected objects are defined and parameterized. The resulting scenario instance can then be used together with the referenced library of building blocks to derive the content of the FOM. The introduction of the layer with the abstract classes also has the advantage that consistent interfaces are made available for e.g. the implementation of the simulation objects' behavior over time. By hiding the simulation-specific implementations in those abstract classes a separation between model and engine is achieved - which is a posed requirement. To fulfill the third basic requirement of also hiding the HLA functionalities, so that the developer does not need to be proficient in the use of the HLA standard itself, an additional library is introduced. This library contains all needed implementations for object management, encoding and decoding HLA messages, time management, federation management, etc. in a way that is generic enough to be used for every federate in the context of a traffic simulating federation. For this purpose, the HLA Ambassador Library (HLAAL) is introduced, which combines the traditional Federate and RTI ambassadors, and enriches them with further generic (de)coding and communication-related functionalities.

The components described above can be seen on the left side of Figure 7. These must now be converted into a runnable simulation environment at the beginning of a simulation run by a kind of model transformation. To accomplish this, a central component is introduced that can read and use the components described above and generate the components needed for the simulation environment based on them. How this conversion is done in detail will be considered in more detail in the next section. This central component should additionally be the only point that provides possibilities for direct user interaction. Thus, the operation of the simulation system can be kept as simple as possible by providing a library and a scenario instance referencing this library as input and the central component handles all following tasks ranging from reading the inputs, transforming the scenario into a set of runnable federates, and starting the federation. Because of this central role this component is called Simulation Manager in the following.

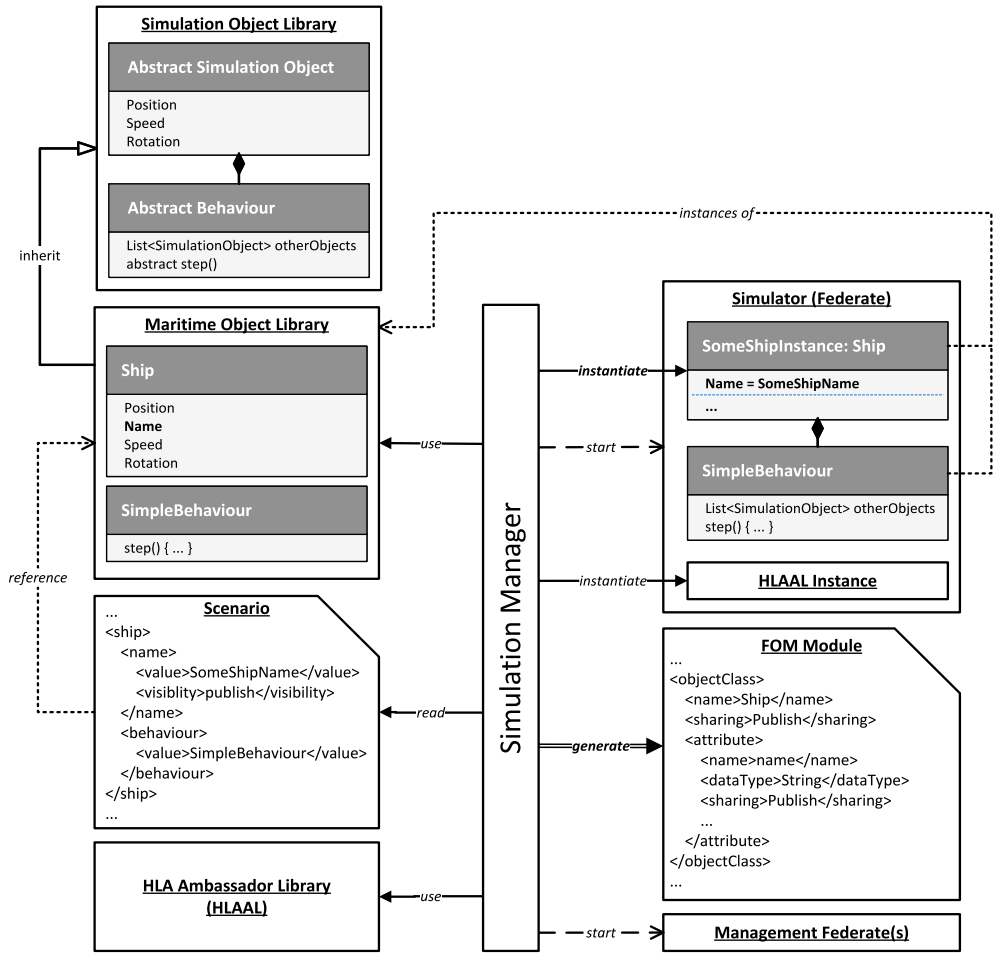

*Figure 7: High-level overview of the scenario to simulation transformation process*

## VII. IMPLEMENTATION

This chapter deals with the implementation of the previously described concept in the form of a first prototype. Challenges to the implementation will be presented, decisions made will be disclosed and justified, and the core elements of the prototype will be examined in more detail. By far not all implementation details are presented and the illustrations are often additionally somewhat abstracted, but the level of detail is sufficient to give a good overview of the way the prototype works. The following chapter then uses a minimal example to show how a scenario can be created and simulated.

Since there are a variety of HLA RTI implementations, recalling that HLA only describes the interfaces and processes but does not provide an implementation, a decision had to be made first on which to use. Since no commercial software was to be purchased for this research work, one selection criterion was that it should be a free open-source implementation. The implementation should also support the latest HLA standard IEEE 1516-2010, also known as HLA evolved, at least in most parts. Out of the four leading HLA RTI implementations according to a recent comparison by Gütlein et al.,[56] there are two non-commercial ones: Portico[57] and CERTI[58]. Portico offers a little more freedom in that it can be used with C++ and Java. In addition, Portico is one of the few implementations that is completely decentralized.[56,59] This decentralization was considered helpful for the fast implementation of a prototype since no central component would have to be set up and the communication among the federates takes place without much configuration effort via multicasts. The RTI component, which has always been illustrated as a central stand-alone component up to now, is merged into the federates, so to speak, by integrating the Portico RTI implementation into each federate, where it performs the tasks of an RTI, such as filtering incoming messages. If the prototype is later further extended, Portico also provides the possibility to use point-to-point connections via a central component called a "WAN router", which is usually the more realistic scenario for productive use.[60] The choice, therefore, fell on Portico. As a result, Java was chosen as the programming language for the implementation of the prototype, since Portico was written in Java and C++ compatibility is only established via wrappers, which require a running JVM container. To avoid this overhead, Java is used directly.

The main challenges for the development were: (1) The conceptualization and prototypical implementation of a data structure for a scenario model library as described in the previous sections; (2) The automated generation of executable sub-simulators from the combination of the scenario instance and the used library. The static content from the scenario instance (XML) and the dynamic content from the library (program code) were to be used; (3) The automated generation of FOM modules for the respective sub-simulators from the given information of the scenario and the library; (4) The automated integration of some kind of Ambassador Library to ensure the uniform interfacing with the local RTI component (LRC); (5) The automated initialization and launch of the Federation and all participating Federates; (6) In addition, it was decided to integrate a specialized observer federate, that subscribes automatically to values previously defined in the scenario instance and makes them available for further use. For the prototype described here, it was decided to send the values to a WebSocket server. However, the Observer-Federate is kept very generic so that the values could also be stored persistently in the future, e.g. in a log file or a database.

A master federate will also be introduced, which will provide the FOM module for all basic simulation functionalities and can be extended in the future to fulfill central management tasks of the simulation system - e.g. stopping the simulation when predefined conditions regarding the simulated objects occur. The usage of Portico, the introduction of an observer federate, and a master federate thus results in the structure of a federation shown in Figure 8.

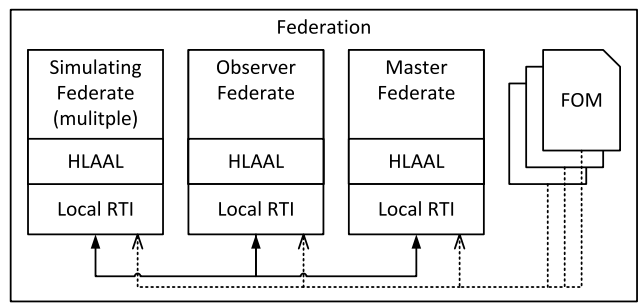

*Figure 8: The resulting structure of an HLA federation based on the prototype implemented here using Portico.*

In order to adopt the overall conceptual structure shown in Figure 7 in the implementation of the prototype, the Java packages were structured accordingly. The resulting package structure can be seen in Figure 9. It can be seen on the left side that the inheritance depth of the library, which serves as a construction kit for the scenario instances, has been extended to three levels. The package *simulation* contains classes, which only represent the basic functionality of a simulation, like the abstract class *SimulationSuperClass* from which all other classes inherit or the abstract class *SimulationObject* which contains basic fields for e.g. the position in the virtual space. Also, the *AbstractBehaviour* is located here, which specifies the implementation of the method *nextStep(double timePassed)*. This is called once per HLA invoked time step during the simulation execution and later contains the user-defined behavioral implementation. A very important class here is also SimulationAttribute<T>, which serves as a wrapper for attributes that can be shared with other federates (publish). In addition to a field for the actual attribute value, the class also contains information that is required to do this, such as the data type of the attribute (Boolean, String, Integer, etc.) and the information on whether this attribute should be shared in the current scenario or not. The package *traffic* imports the package *simulation* and extends the given, mainly abstract, classes by fields and methods, which are necessary for the implementation of traffic. However, this concept of traffic is still unspecific and domain-independent at this point. For example, based on the *SimulationObject*, the class *TrafficParticipant* is introduced, which contains, among other things, the additional fields *SimulationAttribute<Double> speed* and *SimulationAttribute<Double> acceleration*. Together with the *services* package, those two packages form the *base_library* package, which thus represents the concrete implementation of the *Simulation Object Library* from Figure 7. The *services* package offers some functionalities to be used uniformly later, such as converting a scenario instance from a Java object structure to an XML file and vice versa. This functionality is strongly bound to the class structure and is therefore delivered with the library and can be used by a separate application, as the actual simulation system will be.

Building on this generic basis, a traffic domain-specific package can then be defined, as also already indicated in Figure 7. This is also done by inheritance. For example, the *Vessel* class extends the *TrafficParticipant* class with fields representing the Maritime Mobile Service Identity (MMSI) and the current course and draught. The purpose of this structure, as already described in the concept section, is that various domain or use case-specific libraries can be implemented, exported, and made available for use on a consistent and uniform model basis.

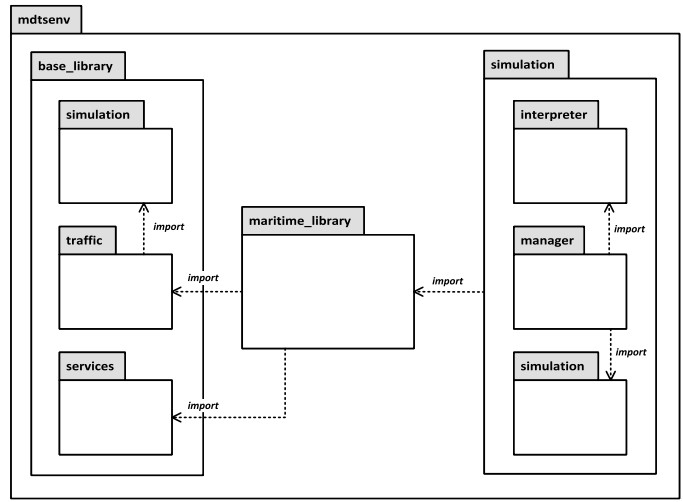

*Figure 9: Package diagram of the implemented Model-driven Traffic Simulation Environment (MDTSEnv)*

The package *simulation* on the other hand contains three sub-packages: *manager*, *interpreter*, and *simulation*. The content of the *manager* package represents the central interface between scenario and simulation (see Figure 7). Located here are mainly functionalities for the import of a scenario instance in the form of an XML file and the control of the program flow during the simulation environment initialization. Functionalities from the *interpreter* package are used, which take care of initializing the scenario contents as an object structure from the combination of the XML file and the given library. In addition, the FOM modules are generated here from this object structure and the XML. The contents of a FOM are first initialized in the form of a Java object structure representing the FOM's content and afterwards this is then converted into XML files according to the standardized format given by HLA's Object Model Template (OMT) Specification and stored temporarily. The reference to this temporary file is then stored in the Java object structure, so that later access (while the Federates join the Federation) is easily possible. During all these transfer and generation processes, all associations between those different kinds of representations of the same objects and attributes, like a FOM path and the corresponding instance of an object, or an HLA compliant FOM and the Java FOM representation, are always stored for easy access in the later course. This is done in a data structuring object called *ReferenceStore* per top-level simulation object (each of those gets its own federate when initializing the simulation environment). The storage of these references has been implemented using the BiMap and Multimap data structures provided by Google's Guava library which extend the Map data structure offered by Java with additional functionality.[61] A *ReferenceStore* uses the fields shown in Listing 1, that can be accessed through public methods for getting, adding, or searching the stored entries like *getFomPathForSimulationObject* and *getSimulationAttributeByUUID*.

```
private ActiveSimulationObject simulationObject;
private String simulationObjectType;
private FOM fom;

private BiMap<String, FOMObjectClass> fomPathToFomObjectClassBiMap;
private BiMap<String, FOMAttribute> fomPathToFomAttributeBiMap;

private Multimap<String, SimulationObject> fomPathToSimulationObjectMap;
private Multimap<String, SimulationAttribute<?>> fomPathToSimulationAttributeMap;

private BiMap<String, SimulationObject> uuidToSimulationObjectBiMap;
private BiMap<String, SimulationAttribute<?>> uuidToSimulationAttributeBiMap;
```

*Listing 1: Field declaration of the ReferenceStore class*

Finally, the sub-package *simulation* contains everything needed for the actual execution of an HLA conforming federation and thus is the implementation of the aforementioned HLAAL. Most of the classes and their interplay can be seen in Figure 10. The central unit is the so-called *InterpretedFederate*, which represents an executable single Federate. Each executable Federate will be assigned a top-level *SimulationObject* from the scenario including the corresponding, previously generated, FOM module and the *ReferenceStore* created during the generating process. Communication with the RTI is handled by an *Ambassador* object and a *DataHandler* object for both the incoming and outgoing directions. The Ambassadors communicate directly with the RTI via standardized interfaces and callbacks. If data is sent (publish) or received (subscribe) about objects or attributes, these again call corresponding methods of the *DataHandlers*. The *DataHandlers* are implemented in such a generic way that only the information from the *ReferenceStore* is needed to encode outgoing data accordingly HLA-compliant or decode received HLA-compliant data to corresponding Java instances of the classes from the library. The receiving process involves storing the object instances that represent the objects published by other Federates in an object instance cache. This set of object instances is then made available to all *Behaviour* implementing classes by injection where it can be accessed in the user implementation of the *nextStep* method. Here it becomes clear why the library was modeled with a multi-layer structure and uniform simple interfaces in mind. This makes it possible to implement behaviors that react to other simulation participants without the developer having to worry about where and how this data comes from.

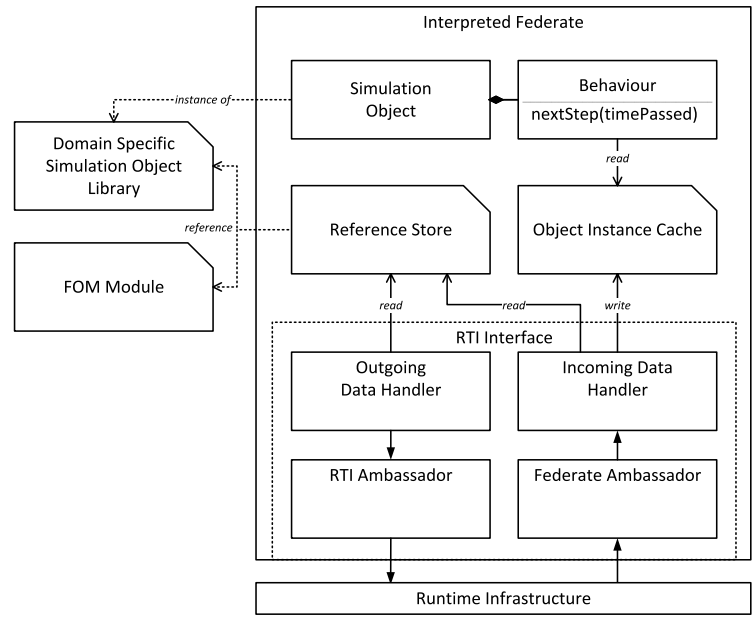

*Figure 10: Overview of the architecture of an initialized InterpretedFederate*

Every concrete behaviour that extends the abstract *Behaviour* and thus implements the interface *IBehaviour* must implement the *nextStep* method as already mentioned. The interface specifies the return type *Map<String, Object>*. A key-value pair of this map contains the internal *ID* of a *SimulationAttribute* and the newly determined value for this attribute, e.g. a new position, a new heading, or a new speed. All attribute value updates returned in this way by all *nextStep* methods executed in the federate are thus first collected and then reflected to the actual attributes identified by the respective *ID*. The central *ReferenceStore* is again used to help with this. This two-part process for updating own attribute values is intended to prevent concurrency problems so that the scenario developer, again, faces less possible complexity.

The entire life cycle of an *InterpretedFederate* is depicted through an UML state machine in Figure 11. It contains all the processes just described above as well as additional simulation management activities and HLA-specific steps. The HLA Development Kit (DKF) framework (see previous section V.C) offers a well-defined behavioral model for managing the life cycle of an HLA Federate[32], which at first glance appears similar. On closer inspection, however, it becomes apparent that the similarity is mainly due to the fact that the HLA standard already sets the basic life cycle phases, such as connecting to an RTI or mechanisms for progression in the simulation time by means of Time Advance Requests (TAR) and Grants (TAG). The concrete federate life cycle conceptualized here differs in that details are included on the management of HLA Handles and the processing of incoming attribute value updates for all kind of objects published by other federates. This is explicitly conceptualized and presented, as this is one of the core features of the approach presented in this paper (cf. *Reference Store* and *Object Instance Cache* in Figure 10).

The *ObserverFederate* relies on a very similar program structure and state flow, but of course, does not have its own *SimulationObject* and thus also works without the execution of behavior. Therefore no outgoing *DataHandler* is needed. Concerning the state flow, the upper sub-state of the running state is therefore also omitted in this case.

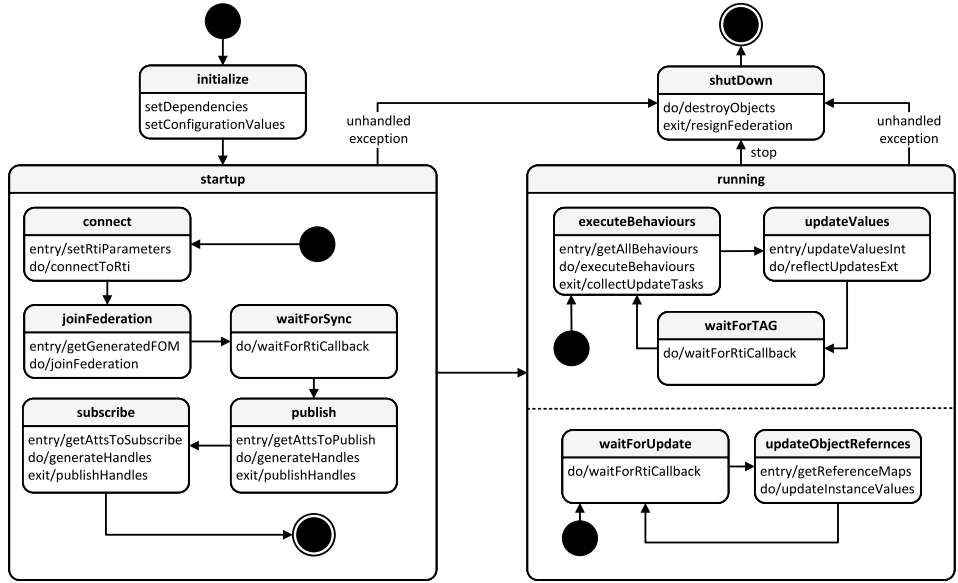

*Figure 11: Overview of the lifecycle of an InterpretedFederate*

## VIII. EXAMPLE OF APPLICATION

In order to verify the functionality and to demonstrate the process of using the implemented prototype, a minimal maritime scenario was realized. Using this example, the process for creating and simulating a scenario is illustrated below. It is assumed that a usage-ready library is given, in the sense of a domain-specific modular set of building blocks, as described in detail in the previous course of this work. The implementation work to create this is therefore not shown in its entirety. However, the procedure for this should have become clear from the description of the package structure earlier.

To demonstrate functionality with minimal scenario complexity, two vessels were placed in open water without any obstacles in the surrounding area. The exemplary made-up scenario is located in the area of the German Bight roughly between Bremerhaven and Wilhelmshaven. The two vessels are roughly based on the characteristics of the container ship *Hamburg Express*[62] and the general cargo ship *Anne-Sofie*[63]. Both should follow a section of a route where Bremerhaven would be the start and the destination the port of Hamburg. In the area north of Wangeroge, however, the specific routes differ to some extent, as do the driven speeds.

In order to represent concrete traffic participating vehicles, the scenario model must be extended as described in the previous chapters. This means that classes have to be defined which inherit from *TrafficParticipant* and describe the specific characteristics of the road users to be simulated. For this concrete scenario this meant that the class *Vessel* defines among others the attribute *vesselName* and is based on the class *TrafficParticipant* of the *base_library*. Inheriting from *Vessel*, the classes *ContainerShip* and *GeneralCargo* provide further properties specific to these types of vessels. The inheritance structure can be seen in Figure 12. Important for the further course is the fact that the attributes highlighted in bold print are distributed over different levels of this inheritance hierarchy and will be important later.

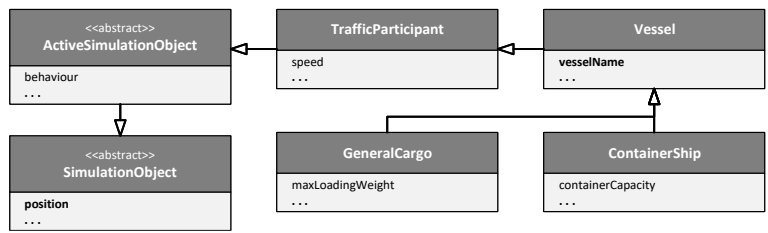

*Figure 12: Class hierarchy across the three levels of the scenario model. The classes Vessel, GeneralCargo and ContainerShip are part of the maritime library used for the application example.*

To define a concrete scenario instance, the possible objects provided by the maritime library are referenced in an XML file and their attributes are provided with concrete values. The desired class, which is to be referenced and later instantiated, is specified via the *type* tag of the *xsi* namespace. The conversion into a Java object structure can then be done later automatically by the interpreter with the help of the Java Architecture for XML Binding (JAXB).

As part of the scenario, the desired observers are also declared and parameterized. In this case, the *ObserverFederate* should subscribe to the four attributes *position*, *vesselName*, *speed,* and *rotation* of all objects of the type *Vessel*. Similarly, it can be defined within a *SimulationObject* to which the corresponding *InterpretedFederate* should subscribe to make the corresponding values available for the contained *Behaviour* implementations. Part of the resulting XML file can be seen in Listing 2. These XML files have currently to be written manually, but some kind of GUI based editor, able to import domain specific libraries and provide the user with an easy way to put together a scenario using the libraries' elements, could be potentially a good tool to ease this heavily manual step in the future.

```xml
<?xml version="1.0" encoding="UTF-8"
    xmlns="http://uol.de/mdts/schema/base"
    xmlns:xsi="http://www.w3.org/2001/XMLSchema-instance"?>

<scenario>
    <library>
        <name>maritime_library</name>
        <version>1.0-SNAPSHOT</version>
    </library>
    <observers>
        <observer>
            <observedClasses>
                <observedClass>
                    <type>vessel</type>
                    <attributes>
                        <attribute>position</attribute>
                        <attribute>vesselName</attribute>
                        <attribute>speed</attribute>
                        <attribute>rotation</attribute>
                    </attributes>
                </observedClass>
            </observedClasses>

            . . .

        </observer>
    </observers>
    <simulationObjects>
        <simulationObject xsi:type="containerShip">
            <behaviour xsi:type="simpleFollowRouteBehaviour" />
            <vesselName>
                <value>Hamburg Express</value>
                <name>vesselName</name>
                <dataType>java.lang.String</dataType>
                <publish>true</publish>
            </vesselName>

            . . .

            <observedClasses>
                <observedClass>
                    <type>containerShip</type>
                    <attributes>
                        <attribute>vesselName</attribute>
                        <attribute>position</attribute>
                    </attributes>
                </observedClass>
            </observedClasses>
        </simulationObject>
        <simulationObject xsi:type="generalCargo">

            . . .

        </simulationObject>
    </simulationObjects>
    <simulationIterations>10000</simulationIterations>
</scenario>
```

*Listing 2: Excerpt from the XML scenario instance describing the example scenario*

The schema referenced in Listing 2 specifies the possible elements of such a scenario XML instance and their structure. Due to the integration of JAXB, the schema can be generated automatically, which means that changes to the basic data model would not require a great amount of manual effort on this end. An excerpt from the schema used here can be seen in Listing 3. It should be noted that *simulationObject*, *behaviour* and *simulationUnit* are declared as abstract. This means that the specific type must still be specified in the scenario XML by using the type attribute of the *xsi* namespace. The concrete implementations of these abstract classes are part of the individually composed domain or use case related libraries. The respective developer is therefore responsible for providing the corresponding schema that builds on this basic schema and describes the concrete characteristics of the contained building block like implementations. An example of this can be seen in Figure 12 and Listing 2: An object of type vessel has a field of type simulationAttribute with the name vesselName. This fact should be represented in a schema belonging to the respective library. In the example from Listing 2, the *simulationObject* is of type *containerShip* and the *behaviour* of type *simpleFollowRouteBehaviour*. The first is indirectly inherited from *activeSimulationObject* (cf. Figure 12) and the second directly from *behaviour*.

```
<?xml version="1.0" encoding="UTF-8" ?>
<xs:schema xmlns:xs="http://www.w3.org/2001/XMLSchema"
           version="1.0.0"
           targetNamespace="http://uol.de/mdts/schema/base">

    <xs:complexType name="scenario">
        <xs:sequence>
            <xs:element minOccurs="0" name="library" type="library"/>
            <xs:element name="simulationIterations" type="xs:int"/>
            <xs:element maxOccurs="unbounded" minOccurs="1" name="simulationObjects" nillable="true" type="simulationObject"/>
            <xs:element maxOccurs="unbounded" minOccurs="0" name="observers" nillable="true" type="observer"/>
        </xs:sequence>
    </xs:complexType>

    <xs:complexType name="library">
        <xs:sequence>
            <xs:element name="name" type="xs:string"/>
            <xs:element name="version" type="xs:string"/>
        </xs:sequence>
    </xs:complexType>

    <xs:complexType name="observer">
        <xs:sequence>
            <xs:element maxOccurs="unbounded" minOccurs="0" name="observedClasses" nillable="true" type="observedClass"/>
            <xs:element minOccurs="0" name="observerWebSocketConfig" type="observerWebSocketConfig"/>
            <xs:element name="timeStepSize" type="xs:double"/>
        </xs:sequence>
    </xs:complexType>

    <xs:complexType name="observedClass">
        <xs:sequence>
            <xs:element maxOccurs="unbounded" minOccurs="0" name="attributes" nillable="true" type="xs:string"/>
            <xs:element name="type" type="xs:string"/>
        </xs:sequence>
    </xs:complexType>

    <xs:complexType abstract="true" name="simulationObject">
        <xs:sequence>
            <xs:element name="formString" type="simulationAttribute"/>
            <xs:element name="physical" type="simulationAttribute"/>
            <xs:element minOccurs="0" name="position" type="simulationAttribute"/>
            <xs:element minOccurs="0" name="rotation" type="simulationAttribute"/>
            <xs:element maxOccurs="unbounded" minOccurs="0" name="observedClasses" nillable="true" type="observedClass"/>
        </xs:sequence>
    </xs:complexType>

    <xs:complexType abstract="true" name="activeSimulationObject">
        <xs:complexContent>
            <xs:extension base="dynamicSimulationObject">
                <xs:sequence>
                    <xs:element name="behaviour" type="behaviour"/>
                    <xs:element name="timeStepSize" type="xs:double"/>
                </xs:sequence>
            </xs:extension>
        </xs:complexContent>
    </xs:complexType>

    <xs:complexType abstract="true" name="behaviour">
    </xs:complexType>

    <xs:complexType name="simulationAttribute">
        <xs:sequence>
            <xs:element name="dataType" type="xs:string"/>
            <xs:element name="name" type="xs:string"/>
            <xs:element name="publish" type="xs:boolean"/>
            <xs:element name="value" type="xs:anyType"/>
        </xs:sequence>
    </xs:complexType>

        . . .

    </xs:complexType>
</xs:schema>
```

*Listing 3: Excerpt from the XML scenario schema referenced in Listing 2*

After starting the implemented prototype simulation system, the scenario instance XML file created as described above can be chosen as direct input file. The appropriate library (currently has to be part of the local java classpath; ideally it will be loaded automatically using the information in the scenario file at some point in the future) is then used to convert the scenario into corresponding Java objects. FOM modules are then generated from the given information for each top-level simulation object. An excerpt from the XML file generated in this way for the *simulationObject* of type *Vessel* shown in Listing 2 can be seen in Listing 4. The inheritance structure from Figure 12 can be recognized here again, which has been converted into an HLA-compliant *objectClass* hierarchy. After all federates have been created and provided with the necessary data and references, they are initialized in an automated way. This means that the master federate gets started and in doing so also creates the federation automatically, afterwards all other generated federates connect to it. As soon as a synchronization point has been reached, the federation is started and time begins to progress, controlled by the RTI. At each time step, each *Behaviour* from the simulation object of an interpreted federate is executed, attributes get updated and the RTI gets informed about those updated values if and only if they were marked as *publish* in the scenario file.

```xml
<?xml version="1.0" encoding="UTF-8"?>
<objectModel xmlns="https://www.sisostds.org/schemas/IEEE1516-2010" xmlns:xsi="https://www.w3.org/2001/XMLSchema-instance"
xsi:schemaLocation="https://www.sisostds.org/schemas/IEEE1516-2010 https://www.sisostds.org/schemas/IEEE1516-DIF-2010.xsd">
    <modelIdentification>
        <name>Vessel--2016892160</name>
        <type>FOM</type>
        . . .
    </modelIdentification>
    <objects>
        <objectClass>
            <name>HLAobjectRoot</name>
            <objectClass>
                <name>SimulationObject</name>
                <sharing>Publish</sharing>
                <attribute>
                    <name>position.latitude</name>
                    <dataType>Double</dataType>
                    <updateType>Unconditional</updateType>
                    <ownership>NoTransfer</ownership>
                    <sharing>Publish</sharing>
                </attribute>
                . . .
                <objectClass>
                    <name>ActiveSimulationObject</name>
                    <sharing>Neither</sharing>
                    <objectClass>
                        <name>TrafficParticipant</name>
                        <sharing>Publish</sharing>
                        <attribute>
                            <name>speed</name>
                            <dataType>Double</dataType>
                            <updateType>Unconditional</updateType>
                            <ownership>NoTransfer</ownership>
                            <sharing>Publish</sharing>
                        </attribute>
                        . . .
                        <objectClass>
                            <name>Vessel</name>
                            <sharing>Publish</sharing>
                            <attribute>
                                <name>vesselName</name>
                                <dataType>String</dataType>
                                <updateType>Unconditional</updateType>
                                <ownership>NoTransfer</ownership>
                                <sharing>Publish</sharing>
                            </attribute>
                            . . .
```

*Listing 4: Excerpt from the automatically generated FOM module XML file describing the data to be shared by the simulation object of type vessel shown in Listing 2.*

To test the availability of published objects and their published attributes of foreign federates, a simple logging output was integrated into the behaviour implementation used. This can be seen in Listing 4. The name of the own simulated ship as well as the class and the current position of the foreign simulation object is written to the console. The console output generated by this code at time step 385 of the simulation is shown in Listing 5.

```java
List<SimulationObject> observedObjects = this.trafficParticipant
    .getObservedObjects()
    .values()
    .stream()
    .findFirst()
    .orElseGet(ArrayList::new);

SimulationObject observedObject = observedObjects.stream().findFirst().orElse(null);

if (observedObject != null) {
    System.out.println("///////////////////");
    System.out.println("// " + vessel.getVesselName().getValue() + ": I CAN SEE!");
    System.out.println("// I HAVE KNOWLEDGE ABOUT THE PARTICIPANT '" + observedObject.getClass() + "'");
    System.out.println("// THE PARTICIPANT CURRENTLY IS AT: " + observedObject.getPosition().getValue().toString());
    System.out.println("///////////////////");
}
```

*Listing 5: Code for test output, which accesses the injected representations of published objects of other Federates.*

```
///////////////////
// Hamburg Express : I CAN SEE!
// I HAVE KNOWLEDGE ABOUT THE PARTICIPANT 'class library.model.maritime.GeneralCargo'
// THE PARTICIPANT CURRENTLY IS AT: "Position":{ "Lat":"53.84009631117777","Lon":"8.115035313513989","Alt":"0.0"}
///////////////////
```

*Listing 6: Test output at simulation time step 1677*

To test and visualize the *ObserverFederate*, a simple NodeJS application providing a WebSocket server, and a simple web application were also implemented. The web application connects to the WebSocket server and displays the available data on top of an OpenStreetMap[64] layer. A screenshot of this can be seen in Figure 13, which was taken at time step 565 of the federation execution. As already mentioned in section 7, the prototypical implementation of the *ObserverFederate* implemented here also sends the received information about the observed objects and attributes to this WebSocket server. This is done in the form of JSON data, which is structured analogously to the Java classes of the used library.

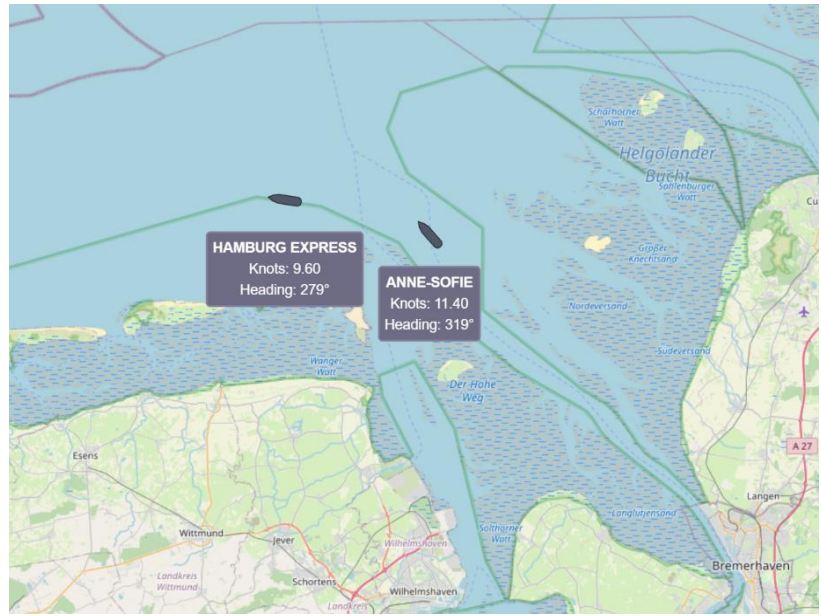

*Figure 13: Screenshot of the web view at simulation time step 1677*
*(Map © OpenStreetMap contributors | www.openstreetmap.org/copyright)*

## IX. DISCUSSION

The exemplary use case shown above is of course far from being a real productive one, but nevertheless, illustrates very well the potential of the approach implemented in this work and how it can be used. It has been shown that it is possible to automate most of the manual steps that were previously necessary for the implementation and execution of HLA-compliant federations. More precisely, using the shown and prototypically implemented approach, no knowledge about HLA itself is necessary anymore: neither by the developer who implements and maintains the domain-specific library nor by the simulation system user who assembles scenarios with the help of the resulting building blocks and then has them simulated. The automatic generation of the Object Model Template (OMT) compliant Federation Object Model (FOM) modules is a particularly important step here.

IEEE Standard 1730-2022 'Recommended Practice for Distributed Simulation Engineering and Execution Process (DSEEP)'[51], that is the successor of the now inactive IEEE Standard 'Recommended Practice for High Level Architecture (HLA) Federation Development and Execution Process (FEDEP)'[40], describes a seven-step process for developing and executing distributed simulation environments. This process reaches from defining the high-level objectives that the simulation environment should fulfill to analyzing the data generated during the simulation runs. The new approach shown and implemented here encompasses a larger number of the sub-steps of this process. Where most related approaches and frameworks only simplify or completely hide two to three steps between step 2 and 4 for the user, the use of the presented approach and prototype can be seen in steps 2 to 6. This makes the overall process more consistent and less complex. **Error! Reference source not found.** directly compares the original DSEEP process and the DSEEP process enriched by the use of the approach presented here. It can be seen immediately that the original steps 3 to 6 are replaced. From the user's point of view, step 5 is completely omitted and steps 3, 4, and 6 are greatly simplified. As mentioned above, the vision is that the developer of the domain-specific library is a different person or group of persons than the one responsible for creating scenarios. Splitting the knowledge and workload like that, also transfers the responsibility for testing to the former. This applies at least to the functionality of the possible simulation elements that are part of the library. Whether a particular scenario run unfolds as planned is still the concern of the scenario creator, but this cannot be equated with a test of whether the federation itself is technically error-free. The start process, originally step 6, is automated to such an extent that the simulation system can be started directly with a scenario instance as only input, which was one of the main requirements to properly support a scenario-based approach in the context of simulative testing. By completely hiding the HLA interface details from the developer and user, the benefits of HLA-based distributed co-simulation can be utilized without the need for familiarization with the complex HLA processes and structures.

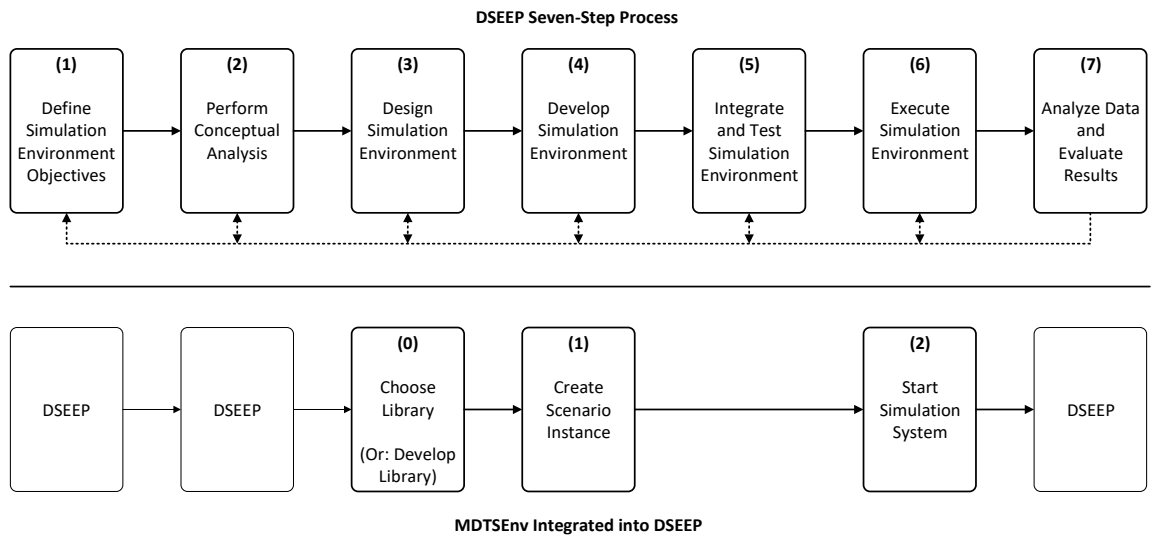

Figure 14: The model-based approach presented and implemented in this work mapped onto the steps of the HLA federation creation workflow according to the DSEEP Framework

Many mechanics of the shown concept and implementation were designed specifically for traffic simulations, respectively for individual traffic participants in cooperation. Within the meta-domain 'traffic' the flexible multi-layer model structure makes it applicable for different concrete traffic domains, but whether the basis of the shown approach can in principle also be used in completely different domains than traffic-related ones would have still to be verified. The connection to a concrete system under test was also considered conceptually in this work but not yet implemented. So, here, too, the chosen path must first prove to be really suitable.

## X. CONCLUSION & FUTURE WORK

In the present work, a first step was taken towards the efficient use of HLA based distributed co-simulation in the context of validation and verification (V&V) of (highly) automated and autonomous traffic systems and subsystems. The necessary smooth transition from a scenario to a running simulation environment, which has to be highly flexible and adaptable to the respective use case, is ensured by the present approach. In the form of a proof of concept, this novel approach was prototypically implemented, and the process was illustrated utilizing an example.

In order to actually productively support the development of corresponding systems and potentially be used in official V&V-based acceptance tests, however, some aspects are still missing due to the prototypical nature of the implementation. Probably the most important point, which was deliberately left out for the most part during the creation of the present concept and implementation, is the connection of the actual system under test (SuT). This can occur in different forms within the context of V&V integrated support for the development of a new system.[17] Therefore, the integration of an SuT in the sense of Model- (MiL), Software- (SiL), Hardware- (HiL) and Vehicle-in-the-Loop (ViL) tests is one of the highest prioritized topics for future work. On the modeling side, a rough concept has already been developed in previous work, but it still needs to be implemented and tested. For ViL tests, where the system under test replaces a complete simulation participant, the idea of some sort of proxy federate is to be further elaborated in the near future. In the implementation of the prototype presented, little value was initially placed on computational performance also. Therefore, there is also a need to catch up here. The current version of the high-level architecture already includes some features that could be beneficial performance-wise. For example, the so-called data distribution management, also known as filtering, can be used to spatially limit which other simulation objects data updates are received by a subscription, which would greatly reduce the amount of network communication. This and other possible techniques will also have to be considered in the future. In its current state, all federates are also run on a single host, which means that the overall simulation is not yet truly distributed, at least not in the physical sense. As already briefly touched on in Section VII, the Runtime Infrastructure (RTI) implementation *Portico* used for the presented implementation offers the possibility of operating federates in a distributed manner even beyond network boundaries, despite its actual decentralized nature. Work has already begun on a way to automate a correspondingly distributed execution of the simulation environment. One of the biggest challenges here is to introduce as few manual steps as possible since one of the main goals of the presented concept is - and will be - to automate the simulative execution of a scenario as much as possible, so that a scenario itself can serve as the only input parameter. A possible platform-independent approach could be the use of the free software for container virtualization *Docker*[65] and the possible generation and distribution of an image per federate. The possibility of physically distributed operation is especially important for the integration of systems as independent federates, as systems to be tested are not necessarily located in the local network. Due to the increasing importance of vehicle-to-vehicle and vehicle-to-anything communication for the functionalities of modern traffic systems, it is also essential to create the possibility that this kind of communication can be directly modeled for and used by the federates. At first glance, Interactions, which are defined by the HLA standard anyway, seem to be a good choice for this purpose. The suitability of these with regard to V2X in the present context of automated FOM generation must therefore be further investigated and a further sub-concept be developed.

To make the creation of the scenario instance XML files less error-prone, it would be useful to have a graphical editor, as already mentioned in section VII. In case this hypothetical editor is not used, however, an XML schema for each individual library should be provided so that users have the possibility to check the correctness of their XML files. The creation of such schema should be relatively easy to automate, since the XML file represents the model structure 1:1 and could therefore be used, for example, with JAXB as seen for the basic scenario scheme in Listing 3, which is already also used to convert the XML scenarios into Java structures.

Another interesting and potentially useful development of the presented assessment would be to establish some compatibility with FMI or individual FMUs to enable hybrid simulations. A lot of work has already been done on this topic, which could possibly be followed up. However, this topic will be put on hold until the approach presented here has been fully proven and exploits the potential of HLA, as this is not a direct requirement for the use case defined here.[66,67]

Despite the multitude of functionalities left to implement and challenges to tackle, the newly created possibility to directly generate simulation models and structures from a model-based scenario instance closes a gap in the traditional use of co-simulations and thus lays the foundation for the efficient use of model- and scenario-based simulation runs in the context of development accompanying V&V processes.